\documentclass[aps,prb,twocolumn,showpacs,groupedaddress]{revtex4}
\usepackage{amsbsy,amssymb,amsmath,bm}
\usepackage{graphicx}

\input{epsf}

\begin{document}

\title {Double resonance response of a superconducting quantum metamaterial: manifestation of  non-classical states of photons}
\author{M. A. Iontsev$^{1}$, S. I. Mukhin$^{1}$ and M. V. Fistul$^{1,2,3}$}
\affiliation {$^{1}$ Theoretical Physics and Quantum Technologies Department,
National University of Science and Technology "MISIS", 119049 Moscow, Russia\\
$^{2}$ Theoretische Physik III, Ruhr-Universit\"at Bochum, D-44801 Bochum, Germany\\
$^{3}$ Russian Quantum Center, 143025 Moscow region, Russia  }

\date{\today}
\begin{abstract}
We report a theoretical study of ac response of superconducting quantum metamaterials (SQMs), i.e. an array of qubits (two-levels system) embedded in the low-dissipative resonator. By making use of a particular example of SQM, namely the array of charge qubits capacitively coupled to the resonator, we obtain a second-order phase transition  between an incoherent (the high-temperature phase) and  coherent (the low-temperatures phase) states of photons. This phase transition in many aspects resembles the paramagnetic-ferromagnetic phase transition. The critical temperature of the phase transition, $T^\star$, is determined by the energy splitting of two-level systems $\delta$, number of qubits in the array $N$, and the strength of the interaction $\eta$ between qubits and photons in the cavity. We obtain that the photon states manifest themselves by resonant drops in the frequency dependent transmission $D(\omega)$ of electromagnetic waves propagating through a transmission line weakly coupled to the SQM. At high temperatures the $D(\omega)$ displays a single resonant drop, and at low temperatures a peculiar \emph{double resonance response} has to be observed. The physical origin of such a resonant splitting is the quantum oscillations between two coherent states of photons of different polarizations.
\end{abstract}

\maketitle

\section{Introduction}
Great attention is devoted to a theoretical and experimental study of novel superconducting quantum metamaterials (SQMs) \cite{SQM1,SQM2,SQM3,SQM4,SQM5,SQM6,SQM7}. The SQMs consist of an array of superconducting qubits, e.g. charge qubits \cite{ChQubit}, flux qubits \cite{Flqubit}, transmons \cite{TRqubit} etc., embedded in low-dissipative resonator. Various macroscopic quantum coherent effects, such as coherent quantum oscillations between two states, microwave induced Rabi oscillations, Ramsey fringes, just to name a few, have been observed in the SQMs. Moreover, since a strong long-range interaction between qubits is provided by exchange of resonators photons, one  can expect a strong variation of the energy spectrum of the SQMs with respect to a set of non-interacting qubits, and therefore, various collective coherent quantum effects in the SQMs. Indeed, in the Ref. \cite{SQM6} instead of a large amount of small different splittings, a single giant splitting has been observed in the spectrum of the SQM, and this effect  indicates the presence of collective  quantum beatings in the SQM.

The various quantum mechanical phenomena manifest themselves by resonant drops in the frequency dependent transmission coefficient $D(\omega)$ of electromagnetic waves  propagating through the transmission line coupled (inductively or capacitively) to the SQM \cite{SQM3,SQM4,SQM5,SQM6}. Such measurement setup is presented schematically in Fig. 1. The theoretical analysis allowing one to express the transmission coefficient $D(\omega)$ in terms of the quantum-mechanical time-dependent correlation function of system of qubits, has been done in Ref. \cite{FistVolkov}.

On the other hand, the interaction of photons with an array of qubits results not only in the change of  qubits spectrum but also it can lead to appearance of  \emph{novel photon states}  in the SQMs.  Indeed,  it was predicted  for the chain of small Josephson junctions \cite{Stroud} and later for the chain of SQUIDS biased in the macroscopic quantum regime, i.e. flux qubits, \cite{FistMukhin} that at low temperatures the coherent state of photons occurs. Moreover, the second-order phase transition between the incoherent state of photons (the high temperature phase) and the coherent state of photons (the low-temperature phase) has been established \cite{FistMukhin}.  However, the physical properties of low-temperature phase have not been studied yet, and the analysis of the transmission wave coefficient $D(\omega)$ for different photonic states has not been carried out.

Therefore, in this Article by making use of a specific SQM, i.e. an array of charge qubits capacitively coupled to the resonator (see, Fig. 1), we provide a complete analysis of both the phase transition  and physical properties of  the photonic states in the low-temperature phase. By making use of a generic study of the transmission coefficient $D(\omega)$ we show that the low temperature phase of photonic states manifests itself by  a peculiar double resonance response. This phenomena has an origin in the quantum beating between the coherent states of photons of different polarizations.

The paper is organized as follows: in Section II we present a particular model, elaborate the Hamiltonian and the effective action of the SQM. In the Sec. III we analyze in detail the phase transition in the photonic states. Especially, we will discuss  the properties of the coherent low-temperature phase. In Sec. IV we apply the generic analysis in order to obtain the transmission coefficient $D(\omega)$ for different photonic states in the SQMs. The Section V provides conclusions.
\begin{figure}[tbp]
\includegraphics[width=4in,angle=0]{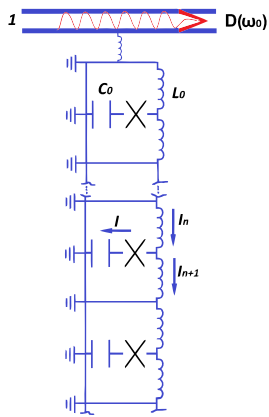}
\caption{ The schematic of the measurements setup: the transmission line inductively coupled to the particular SQM,  i.e. an array of charge qubits embedded in the resonator.}
\end{figure}

\section{Model, Hamiltonian and Effective Action of SQM}
Let us to consider a particular SQM containing the array of $N$ small voltage gated Josephson junctions, i.e. charge qubits. Each qubit is capacitively coupled to the resonator. The schematic of such SQM is presented in Fig. 1. The classical description of the SQM is based on the Lagrangian formalism, and the  Lagrangian of the whole system, i.e. the array of qubits interacting with the resonator, consists of three parts: the photons field,  the array of Josephson junctions and interaction between them:
\begin{equation} \label{Lagrangian}
L=L_{ph}+L_{JJ}+L_{int}.
\end{equation}
The explicit expression for the Lagrangian is derived from the classical equation of motion, where the resonator  is characterized by the time and coordinate dependent charge distribution, $Q(x,t)$, and the dynamics of a single Josephson junctions is described by the time dependent Josephson phase, $\varphi_i(t)$.
By making use of the analysis elaborated in \cite{FistMukhin,Schumeiko,Wallraff} we obtain:
\begin{equation} \label{LagrangianPh}
L_{ph}=m\left[\dot{Q}^2-c_0^2 \left (\frac{\partial Q}{\partial x}\right)^2\right],
\end{equation}
where the effective mass $m=L_0 l/2$, and $L_0$ is the inductance per unit length of the resonator, $l$ is  the total length of the resonator, $c_0$ is the velocity of electromagnetic waves in the resonator.

The Lagrangian of the array of N small Josephson junctions $L_{JJ}$ is written as:
\begin{equation} \label{LagrangianJJ}
L_{JJ}=E_J\sum_i \left \{ \frac{1}{2\omega_p^2}\left [\dot{\varphi}_i+\sqrt{\frac{C_0}{C_J}}2eV_i/\hbar \right ]^2-[1-\cos \varphi_i] \right \},
\end{equation}
where $E_J$  is the Josephson coupling energy, $\omega_p$ is the plasma frequency of the Josephson junction, and $V_i$ are the tunable gate voltages allowing one to control the dynamics of qubits. Here, $C_0$ and $C_J$ are the gate and Josephson junction capacitances, accordingly.

The last part of Lagrangian describes the capacitive interaction between the electromagnetic field and the array of Josephson junctions:
\begin{equation} \label{LagrangianINT}
L_{int}=\frac{\hbar}{2e}\sum_i Q(t,x_i) \dot{\varphi_i}~.
\end{equation}

Next, we greatly simplify a whole problem by taking into account the interaction of arrays of qubits with a \emph{single cavity mode }. In this case the photon mode $Q(x,t)=Q(t)\cos(k_nx)$, where $k_n$ are the wave vectors of cavity modes. Substituting this expresson in Eqs. (\ref{LagrangianPh},\ref{LagrangianINT}) we obtain
\begin{equation} \label{LagrangianPh1}
L_{ph}=\frac{m}{2}(\dot{Q}^2-\omega_0^2 Q^2),
\end{equation}
where $\omega_0=c_0k_n$ is the photon frequency of the resonator. The interacting Lagrangian is written in the following form:
\begin{equation} \label{LagrangianINT1}
L_{int}= \sum_i \tilde{\eta}_iQ(t) \dot{\varphi_i},
\end{equation}
where $\tilde{\eta}_i=\frac{\hbar}{2e}\cos(k_n x_i)$. With an assumption that the size of the array of qubits is much smaller than the resonator size,  all parameters $\tilde{\eta}_i$ are equal to $\tilde{\eta}$.

The equilibrium state of the SQM is described by the partition function $Z$ that can be written through the path integral in the imaginary time representation as
\begin{equation} \label{Zgeneral}
Z=\int D[Q]D[\varphi_i]\exp \left \{-\frac{1}{\hbar}\int_0^{\hbar/k_BT}d\tau L[Q(\tau),\varphi_i(\tau)] \right \}.
\end{equation}
In order to describe the quantum dynamics of small Josephson junctions array interacting with the resonator electromagnetic field, we consider the particular case as all $V_i$ are equal to the same  value of $eV_0=\sqrt{C_J/C_0}(\hbar \omega_p)^2/4E_J$. In this case the quantum dynamics of a single Josephson junction is truncated to the dynamics of a two-level system (TLS) \cite{ChQubit}, i.e. the Hamiltonian $H_{JJ}$ of Josephson junctions array is written as
\begin{equation} \label{HamiltonianArray}
\hat H_{TLS}=\frac{\delta}{2}\sum_i \hat{\sigma}^{(i)}_x,
\end{equation}
where $\delta$ is the splitting between energy levels of the TLS. In the particular case of the array of charge qubits, $\delta=E_J$.
Correspondingly, the interaction of TLSs with the electromagnetic field of the resonator is described by the Hamiltonian, $H_{int}$, as
\begin{equation} \label{HamiltonianINT}
\hat{H}_{int}=\eta Q(t)\sum_i \hat{\sigma}^{(i)}_z,
\end{equation}
where $\eta=\frac{\hbar^2 \omega_p^2}{2E_J}\tilde{\eta}/\hbar$, and $\hat{\sigma}_z$, $\hat{\sigma}_x$ are the corresponding Pauli matrices. In adiabatic regime as the self-frequency of the resonator $\omega_0$ is less than $\delta/\hbar$ we obtain the adiabatic energy levels: $E_{1,2}=\pm \sqrt{\delta^2 +(\eta Q)^2}$ of a single qubit interacting with electromagnetic field of the resonator. By making use of the procedure elaborated in Ref. \cite{FistMukhin} we trace the expression of the partition function $Z$ over the variables, $\varphi_i$, and obtain the effective nonlinear Lagrangian which depends on the photon variable $Q$, only:
$$
Z=\int D[Q]\exp \left \{-\frac{1}{\hbar}\int_0^{\hbar/k_BT}d\tau L_{eff}[Q(\tau)]\right \}~,
$$
\begin{equation} \label{Effective Lagrangian}
L_{eff}=\frac{m}{2}\left [\dot{Q}^2+\omega_0^2 Q^2 \right]-k_BTN \ln \cosh \left [\frac{\sqrt{\delta^2+(\eta Q)^2}}{2k_B T} \right ]
\end{equation}
Thus, one can see that the interaction of photons of resonators with the array of qubits results in the effective nonlinear interaction between  photons.
\section{Phase transition in states of photons}
The  photonic states stabilized  in the SQMs are essentially determined by the type of the effective $Q$-dependent potential
\begin{equation} \label{Potential}
U(Q)=\frac{m}{2}\omega_0^2 Q^2-k_BTN \ln \cosh \left [\frac{\sqrt{\delta^2+(\eta Q)^2}}{2k_B T} \right ]~.
\end{equation}
The potential changes its form at the transition temperature $T^{\star}=\delta \{k_B \ln [(1+\alpha)/(1-\alpha)] \}^{-1}$, where the parameter  $\alpha= \frac{2m \delta \omega_0^2}{ N \eta^2}$.
At high temperatures, i.e. $T >T^{\star}$, the $U(Q)$ has a single minimum at $Q=0$ (see Fig. 2a, blue line). In this case the photon state is the \emph{incoherent one}, and the interaction between photons results in a decrease of the frequency of photons to, $\omega_1=\omega_0 \sqrt{ 1-\frac{1}{\alpha} \tanh{ \frac{\delta}{2k_B T} } }$. For the incoherent state of photons the Kerr type of nonlinearity, $~K Q^4/4$, occurs in the SQM. The Kerr constant $K$ has a following form:
\begin{equation} \label{KerrNonlinear}
K(T)=\frac{N}{\eta^4}{4\delta^3}\left [tanh (x)-\frac{x}{cosh^2(x)} \right]~, ~x=\delta/(2k_BT).
\end{equation}
The Kerr constant is rather small in the limit $T \gg T^{\star}$. The temperature dependence of $K$ is shown in Fig. 2b.

However, at low temperatures $T<T^{\star}$ the effective potential $U(Q)$ has two minima at $Q_{\pm}=\pm \sqrt{\frac{m |\omega_1|^2}{K}}$ separated by the maximum at $Q=0$ (see Fig. 2a, red line). Each minimum corresponds to the \emph{coherent state} of photons. These states are characterized by non-zero values of quantum-mechanical average of charge amplitude, $<Q>=Q_{\pm}$. Frequency of such photon field is also renormalized to, $\omega_2=2 \omega_0 \sqrt{-1+\frac{1}{\alpha}\tanh{\frac{\delta}{2k_B T}}}$. The temperature dependence of the photon frequency in a whole range of temperature is shown in Fig. 2c for different values of parameter $\alpha$.

Moreover, these two coherent states of photons are degenerate states having the same energy but they differ by the polarization of electromagnetic field. The macroscopic quantum tunneling through the barrier (see, Fig. 3a) results in a small splitting $\Delta$ between photonic states macroscopic energy levels,. As a consequence the coherent quantum Rabi oscillations between these macroscopic quantum photon states with the frequency $\omega_R=\Delta/\hbar$ can be established. The frequency of such oscillations is rather small, and it is  obtained in the quasiclassical approximation as
\begin{equation} \label{RabiFrequency}
\omega_R=\Delta/\hbar=\omega_2 \exp \left [ -\frac{2}{3\hbar}\sqrt{2m \lambda}Q_{+}^3 \right ]~.
\end{equation}

\begin{figure}[tbp]
\includegraphics[width=2.8in,angle=0]{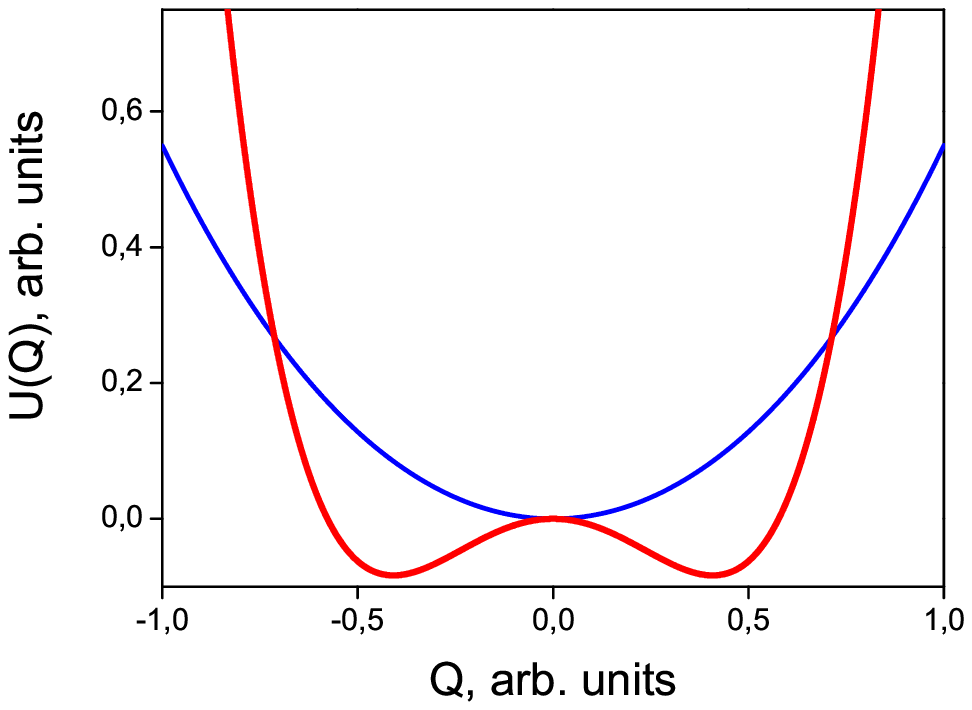}
\includegraphics[width=2.8in,angle=0]{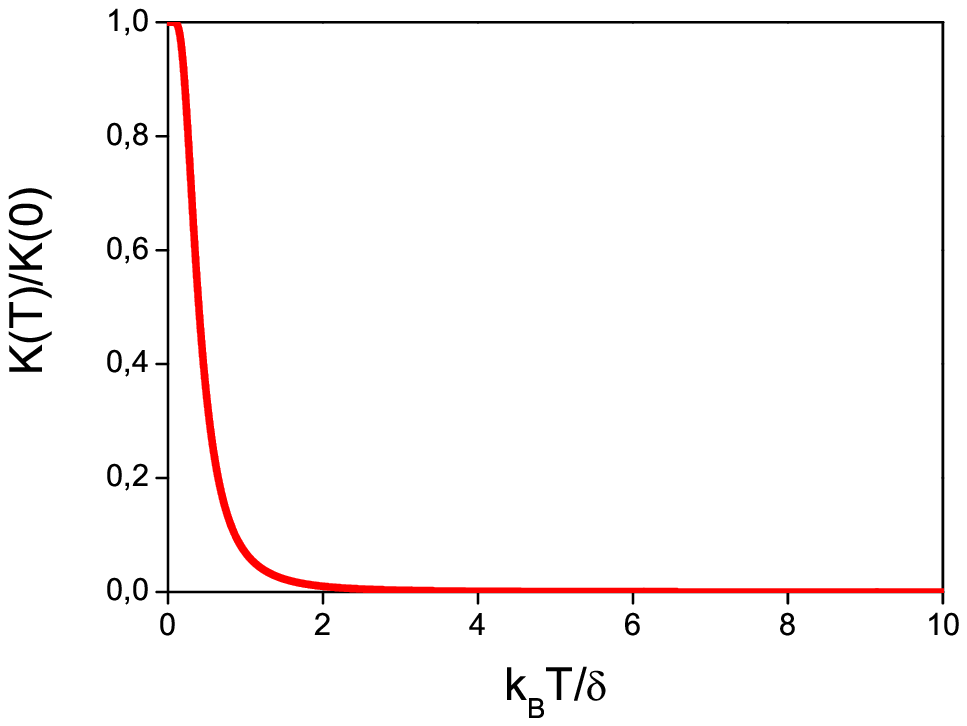}
\includegraphics[width=2.8in,angle=0]{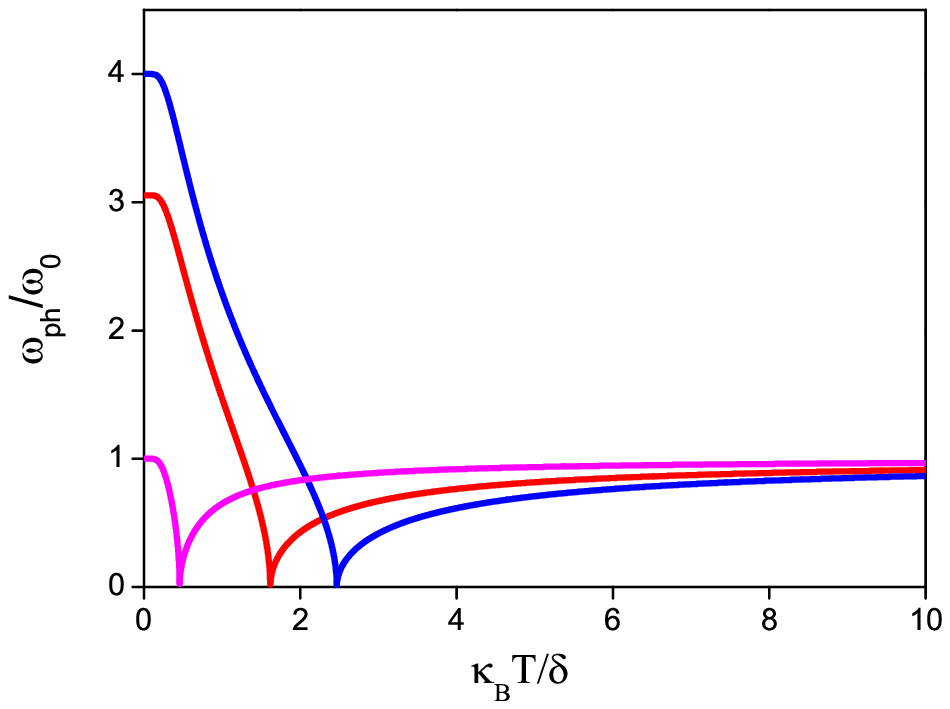}
\caption{a) The effective potential $U(Q)$ describing the interaction of photons in the SQM: high-temperature incoherent phase (blue line, $k_B T=3.3 \delta$) and low temperature coherent phase (red line, $k_B T=0.7\delta$).   b) The temperature dependence of the Kerr nonlinearity parameter $\lambda(T)$. For Figs. 2a(b) the value of parameter $\alpha=0.3$ was used. c) The temperature dependence of the photon frequency $\omega_{ph}$, i.e. $\omega_1$ for $T>T^{\star}$ and  $\omega_2$ for $T<T^{\star}$, for different parameters $\alpha=0.2$ (blue line), $\alpha=0.3$ (red line), and $\alpha=0.8$ (magenta line) }
\end{figure}

\section{AC Response of a quantum metamaterial}
As it was shown in Ref. \cite{FistVolkov} the quantum dynamics of SQMs is directly observed by  measurements of electromagnetic wave (EW)  propagation in the transmission line coupled to the SQM. Such measurement setup is shown in Fig. 1. Similarly, the different states of photons considered in the Sec. III  manifest themselves in the frequency dependent transmission coefficient, $D(\omega)$. Indeed, the EWs propagation in the transmission line is determined by the following equation:
\begin{equation} \label{InhomEquation}
\frac{1}{c_0^2}\frac{\partial^2 q(y,t)}{\partial t^2}-\frac{\partial^2 q}{\partial y^2}=\kappa \delta (y-y_0)Q(y_0,t)~~,
\end{equation}
where $y$ is the coordinate along the transmission line, $q(y,t)$ is the charge distribution in the EW, $\kappa$ is the inductive coupling between the transmission line and the SQM. The Hamiltonian of the SQM interacting with the EWs in the transmission line is written as
\begin{equation} \label{Hamiltonian-InterTransLine}
\hat{H}=\hat{H}_{SQM}-\kappa q(y_0,t)Q(y_0,t)
\end{equation}
The  right-hand part of Eq.(\ref{InhomEquation}) is determined by the quantum-mechanical average of $Q(t)$, i.e.
$$
<Q(t)>=\kappa \int_0^t ds \chi_{QQ}(t-s)q(s),
$$
where $\chi_{QQ}(t)=\frac{i}{\hbar}<[Q(t),Q(0)]>$ is the imaginary  part of the correlation function $C(t)$ \cite{Ingold}. By making use of the Fourier transformation we arrive on the well-known  problem: the propagation of EWs in the 1D channel  in the presence of a single scatterer. Thus, we obtain the transmission coefficient as:
\begin{equation} \label{Transmission coefficient}
D(\omega)=\frac{1}{1+\frac{c\kappa \Im m (\chi_{QQ}) }{\omega}}
\end{equation}
Therefore, the singularities of the $\chi_{QQ}(\omega)$ determine the resonant drops in the $D(\omega)$ dependence. In the high-temperature incoherent photon state the $\chi_{QQ}(\omega)$ is the response function of the harmonic oscillator of the frequency $\omega_1$. By making use of a standard  analysis \cite{Ingold} we obtain
\begin{equation} \label{CorrelFunction-2}
\chi^{incoh}_{QQ}(\omega)=\frac{1}{2m\omega_1}\frac{1}{\omega_1-\omega-i\gamma},
\end{equation}
where $\gamma$ is  the dissipation parameter.
Therefore, the frequency dependence $D(\omega)$ shows a single resonant drop at the frequency$\omega_1$ (see Fig. 4, blue curve) as the incoherent state of photons occurs in the SQM.

The situation drastically changes for the low-temperature phase, where the photonic states have four low-lying coherent states with the energies, $E_{1,2}=\pm \Delta/2$, $E_{3,4}=\hbar\omega_2 \pm \Delta/2$. Moreover, the external EW can excite the transitions between different parity states, i.e. $E_1  \rightarrow E_2$, $E_1 \rightarrow E_4$, and $E_2  \rightarrow E_3$.
By making use of the generic expression \cite{Ingold}
\begin{equation} \label{CorrelFunction-COH}
C(t)=\sum_n \rho_n \sum_m \exp [i(E_n-E_m)t] |<m|Q|n>|^2,
\end{equation}
where $\rho_n$ is the equilibrium density matrix, $<m|Q|n>$ are the matrix elements for the $Q$ operator, the quantum-mechanical  correlation function of the low-temperature photonic state contains three resonant terms:
$$
\chi^{coh}(\omega)= \frac{Q_+^2}{\omega_R-\omega-i\gamma}+
$$
\begin{equation} \label{CorrelFunction-COH2}
+\frac{\hbar}{m\omega (\omega_2+\omega_R-\omega-i\gamma)}+\frac{\hbar}{m\omega (\omega_2-\Omega_R-\omega-i\gamma)}
\end{equation}
It leads to a single drop at low frequencies $\simeq \omega_R$ and double-resonant drop around the photon frequency $\omega_2$ in the frequency dependence of the transmission coefficient $D(\omega)$ (see Fig.4, red line). Therefore, such resonant structure of the frequency dependent transmission coefficient is a fingerprint of macroscopic quantum oscillations between two Bose-condensate of photons.

\begin{figure}[tbp]
\includegraphics[width=2.8in,angle=0]{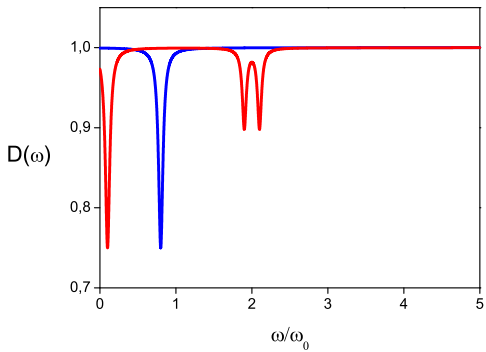}
\caption{The frequency dependent transmission coefficient $D(\omega)$ for the incoherent (blue curve) and coherent (red curve) photon states. The parameters $\omega_1=0.8 \omega_0$ corresponding to the temperature $k_B T=4\delta$, $\omega_2=2\omega_0$ ($k_B T=\delta$ ), $\omega_R=0.1 \omega_0$, and $\gamma=0.03\omega_0$ were used.
}
\end{figure}

\section{Conclusions}
In conclusion we have studied the various equilibrium photonic states occurring  in the SQM, i.e. an array of superconducting qubits embedded in the low-dissipative resonator. We considered the adiabatic non-resonant regime as the photon energy of  resonator $\hbar \omega_0$ is much smaller than the energy splitting of qubits, $\delta$. In this regime we obtained the second-order phase transition in the states of photons. At high temperatures $T>T^\star$ the incoherent state of photons can be realized. In this case the interaction between photons and qubits results in a substantial decrease of the photon frequency as the temperature becomes closer to the transition temperature $T^\star$ (see Fig. 3c). Moreover, the temperature dependent Kerr type of nonlinearity having a quantum origin occurs in the SQM (see Fig. 3b).
At low temperatures ($T<T^\star)$ the coherent states of photons with two different polarizations occur in such a SQM. The frequency of coherent photons increases with temperature (see Fig. 3c). The density of photons in these states is determined by the macroscopic value of $Q_{\pm}$. However, it is most interesting that these two macroscopic coherent states of photons have equal energies, but they are divided by the barrier. Thus, the coherent quantum oscillations of frequency $\omega_R$ between coherent states of photons can be provided by quantum tunneling through the barrier (see Fig. 2a).

By making use of an analysis of the EW propagation in the transmission line coupled to the SQM (see setup in Fig. 1) we obtain that different photon states manifest themselves as resonant drops in the frequency dependent transmission coefficient $D(\omega)$ (see Fig. 4). The incoherent state of photons displays a single drop at $\omega=\omega_1$ but the coherent state of photons has to show three resonant drops: at small frequency $\omega_R$ and double-resonant drop at
frequencies $\omega_2 \pm \omega_R$. A crucial condition to observe these features is a low dissipation in the SQM, i.e. $\gamma < \omega_R$. The observation of such resonant structure in the $D(\omega)$ dependence provides the direct evidence of macroscopic quantum oscillations between two  photonic condensates.

\textbf{Acknowledgments}

We acknowledge the financial support
from the Ministry of Education and Science of Russian Federation in the frame of Increase Competitiveness Program of the NUST MISIS (contracts no. K2-2014-015). MVF acknowledges hospitality of the International Institute of Physics, Natal Brazil where this work has been finished.

{}


\begin{thebibliography}{}

\bibitem{SQM1} A. M. Zagoskin, \emph{Quantum Engineering: Theory and Design of Quantum Coherent Structures.}, Cambridge University Press, Cambridge, 272Ã311 (2011); A. L. Rakhmanov, A. M. Zagoskin, S. Savel'ev, and F. Nori,  Phys Rev B \textbf{77}, 144507 (2008).

\bibitem{SQM2} P. Jung, A. V. Ustinov, and St. M. Anlage, \emph{Progress in Superconducting Metamaterials},  Supercond. Sci. Technol. \textbf{27}, 073001 (2014)
\bibitem{SQM3} Z.-L. Xiang, S. Ashhab, J. Q. You, and F. Nori, Rev. Mod. Phys \textbf{85 }, 623 (2013).

\bibitem{SQM4}J. M. Fink, R. Bianchetti, M. Baur, M. G\"{o}ppl, L. Steffen, S. Filipp, P. J. Leek, A. Blais, and A. Wallraff,  Phys. Rev. Lett \textbf{103}, 083601 (2009).
\bibitem{SQM5}M. Jerger, S. Poletto, P. Macha, U. Huebner, A. Lukashenko, E. Il'ichev, and A. V. Ustinov, Europhys. Lett.\textbf{ 96}, 40012 (2011)
\bibitem{SQM6} P. Macha, Gr. Oelsner, J.-M. Reiner, M. Marthaler, St. Andre, G. Sch\"{o}n, U. Huebner, H.-G. Meyer, E. Il'ichev, and A. V. Ustinov,
Nat. Commun. \textbf{5}, 5146 (2014).
\bibitem{SQM7}D. S. Shapiro, P. Macha, A. N. Rubtsov, A. V. Ustinov, Photonics 2, 449 (2015).


\bibitem{ChQubit} Y. Nakamura, Yu. A. Pashkin and J. S. Tsai,  Nature \textbf{398}, 786 (1999).

\bibitem{Flqubit} I. Chiorescu, Y. Nakamura, C. J. P. M. Harmans, and J. E. Mooij, Science \textbf{299}, 1869 (2003).

\bibitem{TRqubit}J. Koch, T. M. Yu, J. Gambetta, A. A. Houck, D. I. Schuster, J. Majer, Al. Blais, M. H. Devoret, S. M. Girvin, and R. J. Schoelkopf
Phys. Rev. A \textbf{76}, 042319 (2007).

\bibitem{FistVolkov} P. A. Volkov and M. V. Fistul, Phys. Rev. B \textbf{89}, 054507  (2014).




\bibitem{Stroud} J. Kent Harbaugh and D. Stroud, Phys. Rev. B \textbf{61}, 14765 (2000).


\bibitem{FistMukhin}  S. I. Mukhin and M. V. Fistul, Supercond. Sci. and Technology, \textbf{26 }, 084003 (2013).

\bibitem{Schumeiko} M. Wallquist, J. Lantz, V. S. Shumeiko, and G. Wendin,
New J. Phys. \textbf{7}, 178 (2005).

\bibitem{Wallraff} Al. Blais, R.-Sh. Huang, A. Wallraff, S. M. Girvin, and R. J. Schoelkopf,
Phys. Rev. A \textbf{69}, 062320 (2004).

\bibitem{Ingold}G.-L. Ingold, \emph{Path Integrals and Their Application to
Dissipative Quantum Systems}, Lect. Notes Phys. \textbf{611}, 1 (2002).



\end{thebibliography}
\end{document}